\documentstyle[aps,epsf,psfig]{revtex} 
\newcommand{\be}[1]{\begin{equation}\label{#1}}
\newcommand{\ee}{\end{equation}}
\renewcommand{\vec}[1]{{\bf #1}}

\newcommand{\Tr}{\mbox{Tr}}
\newcommand{\N}{\mbox{I}\!\mbox{N}}

\newcommand{\et}{{\em et al }}
\newcommand{\mod}{\mbox{mod}}

\begin{document}
\title{
Unitary stochastic matrix ensembles and spectral statistics
}
\author{ Gregor Tanner} 
\address{
School of Mathematical Sciences
\footnote{e-mail: gregor.tanner@nottingham.ac.uk}\\
Division of Theoretical Mechanics\\
University of Nottingham\\
University Park, Nottingham NG7 2RD, UK
}

\maketitle

\begin{abstract}
We propose to study unitary matrix ensembles defined 
in terms of unitary stochastic transition matrices
associated with Markov processes on graphs.
We argue that the spectral statistics of such
an ensemble (after ensemble averaging) depends 
crucially on the spectral gap between the
leading and subleading eigenvalue of the underlying 
transition matrix.
It is conjectured that unitary stochastic ensembles follow one 
of the three standard ensembles of random matrix theory in the
limit of infinite matrix size $N\to\infty$ if 
the spectral gap of the corresponding transition 
matrices closes slower than $1/N$.
The hypothesis is tested by considering several model systems
ranging from binary graphs to uniformly
and non-uniformly connected star graphs and diffusive networks
in arbitrary dimensions.\\
\noindent
{\normalsize Submitted to {\em Journal of Physics A};\\
Version: 16th March 2001.}
\end{abstract}

\section{Introduction}
\label{sec:sec1}
The study of quantum mechanics on graphs has become an important tool
for investigating the influence of classical dynamics on 
spectra, wavefunctions and transport properties of quantum systems.
Quantum networks have been used with great success to model quantum phenomena 
observed in disordered metals and mesoscopic systems 
(Shapiro 1982, Chalker and Coddington 1988); typical behaviour found
in extended, diffusive systems such as localisation - delocalisation 
transitions (Freche \et 1999), multifractal properties of wavefunctions at 
the transition point (Klesse and Metzler 1995, Huckestein and Klesse 
1999), transport properties (Pascaud and Montambaux 1999, Huckestein 
\et 2000) and the statistical properties of quantum spectra 
(Klesse and Metzler 1997) have been studied on graphs in the limit
of infinite network size.  Recently, Kottos and Smilansky (1997, 1999)  
proposed to study quantum spectra of non-diffusive graphs with only 
relatively few vertices or nodes. This work was motivated by 
understanding how the topology of the graph as well as the boundary conditions 
imposed at the vertices influence the statistical properties of 
the eigenvalue spectrum. A closed expression for the level spacing 
distribution of quantum graphs has recently been given by Barra and 
Gaspard (2000).

The statistical properties of general quantum systems have been studied 
intensively over the last two decades or so.  One finds a strong link 
between spectral statistics of a quantum system and properties of the dynamics
of the underlying classical system. Numerical evidence suggests
that the eigenvalue spectrum of a quantum system whose classical limit 
is chaotic follows one of the three standard ensembles in random matrix 
theory (RMT) (Mehta 1991) after suitable energy averaging. Quantum systems 
whose classical limits show mixed or integrable classical dynamics deviate 
from the RMT results.

In this paper we will make a connection between certain unitary matrix 
ensembles and Markov processes on graphs building upon recent work by 
Tanner (2000) and Pako\'nski \et (2001). It will be 
argued that the statistical properties of an ensemble (after ensemble 
averaging) depend crucially on the spectrum of the transition matrix of the 
corresponding Markov process. Convergence of the ensemble average towards 
the three generic unitary matrix ensembles, the 
circular-unitary, -orthogonal or -symplectic ensemble (CUE, COE or CSE) 
is guaranteed in the limit of infinite matrix size only if correlations in 
the stochastic Markov process decay fast enough in this limit. \\ 

Before introducing these concepts in the following sections, we will 
briefly recapitulate the main ideas behind the quantisation of graphs.  
A quantum graph is essentially a network of vibrating strings fulfilling 
certain boundary conditions at the vertices. (In the usual notation of 
graph theory, we will call a directed bond between a 
vertex $i$ and a vertex $j$ of the graph an edge $(ij)$). Wave propagation
on the graph is written in terms of one-dimensional plane waves moving 
in both directions along undirected edges, and one therefore considers 
{\em undirected} graphs in general.  Following the notation of Kottos and 
Smilansky (1997), a quantum graph 
is characterised by an $N$ dimensional, unitary matrix $\vec{S}_B(k)$ with 
$k$ being the wavenumber and $N$ corresponds to the number of {\em directed} 
edges.  It may be written in the form
\be{S_Smi} \vec{S}_B(k) = \vec{D}(k) \vec{V}(k) ;\ee
here, $\vec{D}$ is a diagonal matrix with diagonal elements 
$d_{ii} = \exp({\rm i} k L_i)$ where $L_i$ is associated with the length of 
the edge $i$. The matrix elements of $\vec{V}$ contain the reflection and 
transmission coefficients for edge - edge transitions at vertices which 
depend on the boundary conditions chosen, see Kottos 
and Smilansky (1997, 1999) as well as Akkermans \et (2000) for details.
The matrix $\vec{S}_B$ is a discrete quantum propagator evolving $N$ 
dimensional complex wave-vectors $\Psi$ between edges according to 
$\Psi_{n+1} = \vec{S}_B(k) \Psi_{n}$. The eigenfrequencies of the system are 
the wavenumbers $k_n$ at which stationary solutions exist, i.e., 
$\vec{S}(k_n)$ has an eigenvalue 1.

When studying the connection between the quantum and classical behaviour on 
graphs, we first have to define what we mean by a `classical' dynamics on a 
finite network and second what constitutes the classical or thermodynamic 
limit when letting the size of the network go to infinity. Defining a 
classical deterministic dynamics on a graph which consists of a finite number 
of vertices acting as branching points of the dynamics does indeed not make 
sense in general (Barra and Gaspard 2001). Instead, one may link the 
probabilistic dynamics of a Markov process defined on the same network 
with the quantum evolution described by the unitary matrix $\vec{S}_B$.
Kottos and Smilansky (1997) suggested considering the Markov process defined 
by the transition matrix $\vec{T}$ with matrix elements given by the 
relation $t_{ij} = |{S_B}_{ij}|^2$.  The stochastic matrix $\vec{T}$ is itself 
a propagator describing the time evolution of a probability distribution 
of particles moving stochastically through the network with transition 
probabilities between adjacent edges given by the matrix elements of 
$\vec{T}$. The degree of chaos found in such a stochastic dynamics is 
characterised by the decay of correlation of an initial probability 
distribution and thus by the spectrum of $\vec{T}$. Note that the classical 
Markov process depends on the structure of the graph as well as on the boundary 
conditions via $\vec{V}$ but not on metric properties of the graph entering 
through the phases $L_i$; this is in contrast to the stochastic dynamics 
introduced by  Barra and Gaspard (2001).  Note 
also, that the transition matrices $\vec{T}$ as defined above describe 
transitions between edges of the graph, not between vertices. 

In the following we will generalise this approach by making a connection 
between arbitrary unitary matrices and their associated stochastic transition 
matrices. In section \ref{sec:sec2}, we will briefly discuss the possible 
Markov processes which may be associated with unitary evolution on finite 
graphs and we will specify an ensemble of unitary matrices linked to a 
given transition matrix.
We will in section \ref{sec:sec3} introduce the spectral form factor which is
the statistical quantity considered throughout the paper,
define a classical limit for networks and formulate a random matrix 
conjecture for unitary matrix ensembles in the limit of infinite 
network size.  Some model systems will be studied in more detail in section 
\ref{sec:sec4}. 

\section{Unitary matrices and associated transition matrices}
\label{sec:sec2}
Any unitary matrix $\vec{U}$ of dimension $N$ can naturally 
be linked to a stochastic transition matrix $\vec{T}$ of a 
finite Markov process with time independent transition probabilities 
by making the following connection between matrix elements of $\vec{U}$ and 
$\vec{T}$, i.e.\
\be{quan_class}    
u_{ij} = r_{ij} e^{i\phi_{ij}}  
\longrightarrow t_{ij} = |u_{ij}|^2 = r_{ij}^2\; .
\ee
The matrix $\vec{T}$ is clearly a stochastic matrix due to the unitarity of 
$\vec{U}$, that is, $\vec{T}$ fulfills
\begin{eqnarray} \label{stoch} 
\nonumber
t_{ij} &\ge& 0\; \quad \mbox{for all}\; i,j = 1, \ldots N\\
\sum_{j=1}^N t_{ij} &=& 1\;\quad \mbox{for all}\;i = 1, \ldots N \, .
\end{eqnarray}
This implies that 1 is an eigenvalue of $\vec{T}$, all other eigenvalues 
have modulus less or equal 1.
The matrix elements $t_{ij}$ can be interpreted as probabilities 
for making a transition from a vertex $i$ to a vertex $j$ on a graph of $N$ 
vertices.  The structure of the corresponding graph is explicitly defined 
through $\vec{T}$ and thus $\vec{U}$, i.e., an edge $(ij)$ exists if and only 
if $t_{ij} > 0$. The connection (\ref{quan_class}) provides therefore a natural 
link between a unitary 'quantum' evolution and a stochastic `classical' 
dynamics in the form
\begin{eqnarray} 
\nonumber
\mbox{quantum}\quad\;&\qquad \qquad&\quad\; \mbox{classical}\\
\Psi_{n+1} = \vec{U}^{\dagger} \Psi_n    && p_{n+1} = \vec{T}^{\dagger} p_n \,.
\end{eqnarray}
Here, $\Psi_n$ is a complex $N$ dimensional 
wave-vector propagating through the network and $p_n$ is a 
probability distribution; its $i$-th component corresponds to the probability 
finding a particle at vertex $i$ at time $n$ after having wandered 
stochastically through the network starting with a probability distribution 
$p_0$ at time $n=0$ (Berman and Plemmons 1979). Note that, in 
contrast to quantum graphs briefly described in section 
\ref{sec:sec1}, the graphs defined through general unitary matrices may be 
directed. The dynamics on the graph is now 
defined with respect to the vertices of the graph, not with respect to the 
edges. 

It is conjectured that statistical properties of generic quantum spectra are 
strongly influenced by how fast correlations in the corresponding classical 
dynamics decay (Bohigas \et 1984, Berry 1985). We may thus expect that 
the statistical properties of the spectrum of a unitary matrix $\vec{U}$ are
linked to the properties of the stochastic dynamics generated by the associated
transition matrix $\vec{T}$. Before discussing this further in section 
\ref{sec:sec3}, we will explore the 
link between unitary and stochastic matrices in more detail. 

\subsection{Unitary stochastic matrices}
Not every stochastic matrix $\vec{T}$ fulfilling the conditions (\ref{stoch})
can be associated with a unitary matrix as defined in (\ref{quan_class}).
Unitarity indeed requires that besides (\ref{stoch}), also the condition 
\be{doubly_stoch}
\sum_{i=1}^N t_{ij} = 1 \; \quad \mbox{for all}\; j 
\ee
holds. That is, $\vec{T}$ must be {\em doubly stochastic} 
(Marshall and Olkin 1979).
One deduces immediately that the vector $\tilde{p}$ with components
$\tilde{p}_i = 1/N$ for all $i=1,\ldots N$ is a stationary state, i.e., 
it is the left eigenvector of $\vec{T}$ with eigenvalue 1. 
(It is of course also right eigenvalue of $\vec{T}$ as for all stochastic 
matrices.) This in turn implies, that the Markov process is ergodic 
or irreducible (Berman and Plemmons 1979) if the graph is connected, that is, 
if it is not possible to decompose the underlying graph into disconnected 
subgraphs. A general stochastic matrix is called primitive  
if $\vec{T}^{k}$ is positive for some $k\ge 0$, that is 
$t^{(k)}_{ij} > 0$ for all $i$ and $j$ (Berman and Plemmons 1979). 
This implies that the spectrum of $\vec{T}$ given in terms of the eigenvalues 
$\{\Lambda_0, \ldots, \Lambda_{N-1}\}$ with $|\Lambda_i| \le |\Lambda_j|$ 
for $ i > j$ or in terms of the eigenexponents 
$\{ \lambda_i = -\log|\Lambda_i| \}$ 
has a finite gap between the leading exponent $\lambda_0 = 0$ and the 
next to leading exponents, that is, 
\be{spectrum}
\Delta = \lambda_1 - \lambda_0 = \lambda_1 > 0 \; .
\ee
A doubly stochastic matrix is primitive if the graphs corresponding to 
$\vec{T}^n$ are connected for all $n$. A finite gap in the spectrum means that 
initial probability distributions $p_0$ on the network decay exponentially 
towards the equilibrium distribution $\tilde{p}$ with decay rate 
$\ge \lambda_1$.

Without going further into the theory of doubly-stochastic
matrices 
we note that a doubly-stochastic matrix can be written in terms of $(N-1)^2$ 
independent parameters, say the matrix elements in the 
first $N-1$ rows and columns. These matrix elements are constrained by the 
inequalities
\begin{eqnarray} \label{polytop}
\sum_{i=1}^{N-1} t_{ij} &\le& 1 \quad \mbox{for all}\; j; \qquad 
\sum_{j=1}^{N-1} t_{ij} \le 1 \quad \mbox{for all}\; i; \\
\nonumber
\sum_{i,j=1}^{N-1} t_{ij} &\ge& N-2\,; \qquad \qquad t_{ij} \ge 0 
\quad \mbox{for all}\; i, j 
\end{eqnarray}
and thus correspond to a finite domain in the $(N-1)^2$ dimensional
parameter space.

Not every doubly-stochastic matrix can be associated with a unitary matrix 
as defined in (\ref{quan_class}). 
The rows and columns of a unitary matrix have to obey orthogonality 
conditions which impose further restrictions on the 
matrix elements $t_{ij}$.  One therefore defines the 
subset of doubly-stochastic matrices $\vec{T}$ which fulfill $t_{ij} = 
|u_{ij}|^2$ for some unitary matrix $\vec{U}$ as {\em unitary-stochastic}
transition matrices (Marshall and Olkin 1979). The dimension of the 
parameter space for unitarity-stochastic matrices is $(N-1)^2$ as for 
doubly-stochastic matrices, the parameter space covered 
by unitary-stochastic matrices is, however, in general smaller than the domain
specified in (\ref{polytop}). To get precise bounds for the possible parameters 
for unitary-stochastic matrices is a non-trivial problem in general 
and is beyond the scope of this article, see Pako\'nski \et (2001) for details. 

\subsection{Unitary stochastic ensembles}
Next we will focus on the space of unitary matrices related to a 
given unitary-stochastic matrix $\vec{T}$. This space, together with
a probability measure specified later, forms an ensemble which we will call
a {\em unitary stochastic ensemble} $U(N,\vec{T})$. These ensembles have a
surprisingly simple structure. The number of independent parameters determining 
a unitary matrix uniquely is $N^2$. Of these $(N-1)^2$ parameters are fixed 
by the unitary-stochastic matrix $\vec{T}$, namely the amplitudes 
$r_{ij} = \sqrt{t_{ij}}$. After decomposing $\vec{U} \in U(N,\vec{T})$ in 
the form
\begin{eqnarray} \label{decomp}  
 \vec{U} = \vec{D}_1 \tilde{\vec{U}} \vec{D_2}
         = \left(\begin{array}{cccc} e^{{\rm i} \varphi_1}&&&\\ 
                                  & e^{{\rm i} \varphi_2}&&\\ 
                                  & &\ddots&\\ 
                                  & &&e^{{\rm i} \varphi_N} 
             \end{array}\right)
            \left( \begin{array}{cccc}  r_{11}&r_{12}&\ldots&r_{1N}\\ 
                                  r_{21}&&&\\ 
                                  \vdots&&\vec{W} &\\ 
                                  r_{N1}&&& 
             \end{array}\right)
            \left(\begin{array}{cccc}              1&&&\\ 
                                 & e^{{\rm i} \varphi_{N+1}}&&\\ 
                                 & &\ddots&\\ 
                                 & &&e^{{\rm i} \varphi_{2N-1}} 
            \end{array}\right) ,
\end{eqnarray} 
one finds that the remaining $2N-1$ independent parameters are the phases 
$\varphi_i$ which can take any values in $[0, 2\pi]$. The building block
of the ensemble is the unitary matrix $\tilde{\vec{U}}$ which
has been chosen here to have real, positive matrix elements in the first row
and column. It is connected to the transition matrix $\vec{T}$ via the relation
$|\tilde{u}_{ij}| = r_{ij}$. The phases of $\tilde{\vec{U}}$ are all contained 
in the $(N-1)$ dimensional complex matrix $\vec{W}$ and are fixed by the 
$N (N-1)$ orthogonality conditions between the rows (or columns) of 
$\tilde{\vec{U}}$.
This set of equations has a discrete set of solutions for generic unitary 
stochastic matrices and $N\ge 3$ (Pako\'nski \et 
2001), that is $\vec{W}$ and thus $\tilde{\vec{U}}$
is not uniquely determined by $\vec{T}$. To find all possible solutions for 
$\tilde{\vec{U}}$ is a non-trivial problem in general and is linked to the 
problem of finding the maximally available parameter space for unitary 
stochastic matrices of a given dimension.

The ensemble $U(N,\vec{T})$ as a whole is parameterised by $2N-1$ phases 
$\varphi_i$ at most and has, for fixed $\tilde{\vec{U}}$, the topology of a 
$2N-1$ dimensional torus. The symmetry 
properties of the ensemble are essentially given by the symmetries of 
$\tilde{\vec{U}}$, we expect in particular time reversal symmetry if 
$\tilde{\vec{U}}$ and thus $\vec{T}$ are symmetric. The non-uniqueness  
will play a role only if the possible solutions $\tilde{\vec{U}}$
belong to different symmetry classes. Assuming that this is not the case,
we may treat ensembles for fixed $\vec{T}$ but different $\tilde{\vec{U}}$
as equivalent and we therefore disregard the $\tilde{\vec{U}}$-dependence in 
what follows.

Taking the trivial probability measure on the parameter space $\varphi_1, 
\ldots, \varphi_{2N-1}$, we can perform the ensemble average of a 
function $f(\vec{U})$ by straightforward integration over the angles 
$\varphi$, that is
\be{ava_1}
<f>_{U(N,\vec{T})} = \frac{1}{(2\pi)^{2N-1}} 
\int_0^{2 \pi} d\varphi_1 \ldots \int_0^{2 \pi} d\varphi_{2N-1}  
f(\varphi_1, \ldots, \varphi_{2N-1}; \vec{T})\; .
\ee
The multiple-integral can be reduced to an integral over only the first 
$N$ phases if $f$ depends on the eigenvalues of $\vec{U}$ only. The 
average (\ref{ava_1}) may be written in terms of a `time' 
average over an ergodic path on the torus. After choosing $2 N - 1$ 
rationally independent but otherwise arbitrary length segments 
$L_1, \ldots L_{2N-1}$, one defines the trajectory 
\[(\varphi_1,\ldots,\varphi_{2N-1})(k)= 
(k L_1,\ldots, k L_{2N-1})\,\mbox{mod} 2\pi\]
which covers the torus uniformly when letting the fictitious time $k$ go to 
infinity.  The average is now taken over the one-dimensional parameter family 
\be{U_k} \vec{U}(k) = \vec{D}_1(k) \tilde{\vec{U}} \vec{D}_2(k)\ee
where the $k$-dependence in $\vec{D}_1$ and $ \vec{D}_2$ enters through the 
replacement $\varphi_i = k L_i$ in (\ref{decomp}). This parameterisation is a 
generalisation of the product form in Eqn.\ (\ref{S_Smi}). The average can
now be written as
\be{ava_2}
<f>_{U(N,\vec{T})} = \lim_{k_0\to\infty} \frac{1}{k_0} 
 \int_0^{k_0}  dk\; f(\vec{U}(k)) \;.
\ee
The importance of choosing rationally independent lengths segments $L_i$, 
also stressed by Kottos and Smilansky (1997), becomes obvious. For 
rationally dependent $L_i$'s only a lower dimensional subspace of the 
full parameter space is covered in (\ref{ava_2}) which may lead to 
averages different from the full ensemble average (\ref{ava_1}). 

\section{Spectral statistics for unitary-stochastic ensembles}
\label{sec:sec3}
So far we proposed to divide the unitary group into 
unitary-stochastic ensembles (USE) which are defined explicitly through 
unitary-stochastic matrices $\vec{T}$. We will argue now that the 
spectral statistics of unitary matrices forming a USE depend strongly on 
the eigenvalues of $\vec{T}$. 

\subsection{The spectral form factor}
In the following we identify the spectrum of a unitary matrix $\vec{U}$ of 
dimension $N$ with the set of eigenphases 
$\{\theta_1, \ldots, \theta_N\}$ of $\vec{U}$. 
The statistical measure used is the so called 
spectral from factor, the Fourier-transform of the spectral 
2-point correlation function
\be{two-point1}
R_2(x) = 
\frac{1}{\overline{d}^2} 
<d(\theta) d(\theta + x/\overline{d})>_{U(N,\vec{T}), \theta} \, .
\ee
Here, $d(\theta,N) = \sum_{i=1}^N \delta(\theta - \theta_i)$ denotes the
density of states and the mean density $\overline{d}$ is given by 
$\overline{d} = N/2\pi$ (see e.g.\ Tanner (1999)). The average is taken over 
the angle $\theta$ and a USE.  After averaging out the $\theta$ dependence,
one recovers the Fourier coefficients in terms of the traces of $\vec{U}$, 
i.e., one obtains for the form factor 
\be{K1}
K(\tau) = <\frac{1}{N} |\Tr \vec{U}^{N\tau}|^2 >_{U(N,\vec{T})}
\ee
with $\tau = n/N$ and the average is taken over a USE. The traces of 
$\vec{U}^n$ can be written as (Kottos and Smilansky 1997)
\be{Tr-sc}
\Tr \vec{U}^n = \sum_{p}^{(n)} A_{p} e^{{\rm i} L_p} \, 
\ee
where the summation is over all periodic or closed paths of length $n$ on the 
graph. Characterising a given periodic path by its vertex code 
$(v_1, v_2 \ldots v_n)$, $v_i \in \{1,2,\ldots N\}$ with $(v_i v_{i+1})$ being
allowed transitions between vertices, one obtains, following the notation in 
(\ref{quan_class}),    
\be{weights} 
L_p = \sum_{i=1}^n \phi_{v_iv_{i+1}}, \qquad
A_p = \prod_{i=1}^n r_{v_iv_{i+1}}.
\ee
The form factor can thus be written as a double sum over periodic paths 
on the graph
\begin{eqnarray} \label{k_po1}
K(\tau) &=& <\frac{1}{N} \sum^{(n)}_{p,p'} A_p A_{p'} 
e^{{\rm i} (L_p - L_{p'})} >_{U(N,\vec{T})}\\
\label{k_po2}
 &\approx& g \frac{n}{N} \Tr \vec{T}^n  + <\sum^{(n)}_{p \ne p'} A_p A_{p'} 
e^{{\rm i} (L_p - L_{p'})} >_{U(N,\vec{T})}\,.
\end{eqnarray}
The first term in (\ref{k_po2}) is the so-called diagonal term (Berry 1985).
It stems from periodic orbit pairs ($p, p'$) related through cyclic 
permutations of the vertex symbol code, that is, of orbits with vertex codes
$(v_1, v_2 \ldots v_n)$, $(v_2, v_3 \ldots v_n,v_1), 
 \ldots, (v_n, v_1 \ldots v_{n-2}, v_{n-1})$; there are typically $n$ orbits 
related by cyclic permutations and all these orbits have the same amplitude 
$A$ and phase $L$. The corresponding periodic orbit pair contributions
in (\ref{k_po1}) are thus equal to $A_p^2$ which is the classical probability 
for following the given cycle for one period.
Additional periodic orbit degeneracies may occur due to symmetries. For time 
reversal symmetric dynamics, for example, periodic cycles with symbol code 
$(v_1, v_2 \ldots v_n)$ and $(v_n, v_{n-1} \ldots v_1)$ have identical phases 
and amplitudes. This leads to an additional symmetry degeneracy factor
$g$ in (\ref{k_po2}) which is one for non-time reversal symmetric dynamics 
and two for time reversal symmetric dynamics, for example.

The diagonal term constitutes the important connection between 
the form factor and the stochastic transition matrix $\vec{T}$. 
The second term in (\ref{k_po2})
is a double sum over the remaining periodic orbit pairs. 
Contributions to this term which survive the ensemble average can be 
formulated in terms of periodic orbit degeneracy classes and are due to
phase correlations imposed by unitarity conditions (Berkolaiko and Keating 
1999, Tanner 2000). These contributions are negligible in the limit 
$\tau \to 0$ after ensemble averaging, but are vital to reproduce the form 
factor for finite $\tau$ values.

Working in the diagonal approximation valid in the asymptotic regime 
$n \to \infty$ and $n/N = \tau \to 0$, one obtains
\be{D_approx}
K(\tau)  \approx g \, \tau \, \Tr \vec{T}^n  = 
g\, n\, \overline{P}(n)
\ee
where we introduce the mean return probability per vertex 
(Argaman \et 1993)
\be{RP_def} 
\overline{P}(n) = \frac{1}{N}\Tr\vec{T}^n \, .
\ee
We will argue that the spectrum of $\vec{T}$ determines whether or not the
statistical behaviour of a unitary-stochastic ensembles $U(N,\vec{T})$ 
follows RMT in the classical limit $N\to \infty$. Before doing
so we have to specify more precisely what we mean by the classical or
thermodynamic limit of a stochastic dynamics on a finite graph. 

\subsection{The classical limit and a random matrix conjecture for USE's}
\label{subsec:thermo} 
In what follows we will define the classical limit of a
family of stochastic Markov processes when letting the number of vertices
and thus the dimension of $\vec{T}$ go to infinity.
We thereby distinguish between {\em finite} systems on the one hand and 
{\em extended} systems on the other. 
A series of Markov processes approximating the Perron-Frobenius operator of
a deterministic system acting on a bounded domain 
is a typical example of convergence to a finite classical system. The piecewise 
linear maps on the unit interval considered by Pako\'nski \et
(2001) are particularly simple examples where the leading eigenvalues of the
Perron-Frobenius operator are recovered already by finite transition matrices. 
Transition matrices with increasing 
dimension $N$ resolving the phase space dynamics on finer and finer scales 
are necessary to capture more and more details of the classical dynamics for
generic maps. We will distinguish these types of systems from extended 
systems consisting of networks of connected, equivalent subsystems as for
example the lattices shown in Fig.\ \ref{Fig:diffgraph}. Appropriate 
rescaling with respect to the system size is necessary here to 
define useful quantities describing the dynamical behaviour per 'unit cell'.

To make the notion of a classical limit precise, we will adopt the following
definition in what follows: consider a series of Markov processes
given in terms of transition matrices $\{\vec{T}_i,\, i=1,
\ldots,\infty\}$ with $N_i=\mbox{dim}\vec{T}_i < N_j=\mbox{dim}\vec{T}_j$ for
$i<j$. We will say that such a series has a well defined classical limit 
corresponding to 
\begin{itemize}
\item a finite classical system if the integrated return probability 
$IP_i(n) = \Tr \vec{T}_i^n$ converges uniformly to a limit function
$IP_{cl}(n)$ in the limit $N_i \to \infty$;
\item an extended classical system if the mean return probability per 
vertex 
$\overline{P}_i(n) = \Tr \vec{T}_i^n/N_i$ converges uniformly 
to a limit function $\overline{P}_{cl}(n)$ in the limit $N_i \to \infty$.
\end{itemize}

The semiclassical limit for a family of unitary stochastic ensembles is
then defined via a family of unitary-stochastic transition 
matrices $\{\vec{T}_i\} $ with well defined classical limit in the 
sense above. 

We are now able to formulate a random matrix conjecture for
unitary stochastic ensembles in terms of the spectral gap $\Delta$
similar to the Bohigas-Giannoni-Schmit conjecture
for general quantum systems (Bohigas \et 1984). We propose
\footnote{The condition that the family $\{\vec{T}_i\}$ must have a classical
limit can be relaxed. We indeed expect that USE's corresponding to an arbitrary
series of transition matrices $\{\vec{T}_i\}$ with 
$\Delta_i N_i/N_i^{\delta}$ bounded from below by a positive constant
follow RMT in the limit $N_i \to \infty$. The limit (\ref{conj}) is then, 
however, not defined in general and controlling the bound is difficult in 
practice. The series is in addition  not linked to any specific 
dynamical system.}:\\

\noindent
{\em The spectral 
statistics of a family of unitary stochastic ensembles 
$\{U(N_i,\vec{T}_i)\}$ with associated 
transition matrices $\vec{T}_i$ having a well defined 
classical limit follows one of the three random matrix ensembles 
CUE, COE or CSE in the semiclassical limit if the spectral gap
$\Delta_i = \lambda^{(i)}_1 - \lambda^{(i)}_0 = \lambda^{(i)}_1$ of $T_i$ 
decreases slower than $1/N_i$ in the classical limit, or more precisely, that 
there exists a $\delta_0 > 0$ such that
\be{conj} 
\lim_{i\to\infty} \frac{\Delta_i N_i}{N_i^{\delta}} \ge c >0 
\qquad \mbox{for all} \quad 0<\delta\le \delta_0 .
\ee
}

The conjecture implies that USE's associated with primitive transition 
matrices in the classical limit, that is, matrices with a non-vanishing 
spectral gap and exponential decay of correlation, follow RMT-statistics. 
More important is, that the bound (\ref{conj}) does not exclude
RMT-statistics for classical dynamics with algebraic decay of correlation. 
We also emphasis that it is the spectral gap which is the crucial quantity 
in the conjecture. No reference is made
to the Kolmogorov-Sinai entropy or similar measures of chaos. We will
present systems with positive KS-entropy not following RMT-statistics in the
classical limit in section \ref{sec:sec4}.

The $1/N$ - threshold condition in (\ref{conj}) is a 
consequence of the $\tau = n/N$ scaling; rewriting (\ref{D_approx}) in the 
form
\[ 
| K(\tau) - g \tau| \approx \left|g \tau \sum_{i=1}^{N-1}\Lambda_i^n\right| \le 
g \tau (N-1) e^{-\lambda_1 N \tau}\, ,
\] 
the condition (\ref{conj}) implies that the right hand side vanishes for fixed 
$\tau$ and $N\to\infty$. Much less clear is, however, why the purely
classical condition (\ref{conj}) implies the RMT result for $K(\tau)$ 
for all $\tau$ as conjectured here. This problem lies at the heart of many 
studies conducted in the recent past (Kottos and Smilansky 1999, 
Berkolaiko and Keating 1999, Schanz and Smilansky 1999, 2000, Tanner 2000) 
and will not be addressed further.
We will instead consider in the next section a few model systems with 
spectral gaps both below, on and above the critical threshold and will give
numerical results showing that the threshold condition is indeed vital for 
spectral statistics. 

\section{Numerical results}
\label{sec:sec4}
\subsection{Unitary stochastic ensembles with non-vanishing spectral gap}
\label{subsec:fingap}
We will first discuss families of USE's with transition matrices
having a non-vanishing spectral gap in the classical limit. We expect these 
ensembles to follow RMT-statistics for $N\to\infty$. Two specific examples 
are considered: binary graphs with sparsely filled transition matrices and 
a finite spectral gap and fully connected graphs with uniform transition 
amplitudes having an infinitely large spectral gap. We find in both 
cases power law convergence of the form factor to one of the three RMT - 
ensembles.
\paragraph{Binary graphs.}
We consider a special class of binary graphs with transition matrices 
\be{tran_bin} t_{ij} = \left\{ \begin{array}{ll} 
\frac{1}{2}  &\quad \mbox{if}\quad  j=2i \;\mod N \quad \mbox{or} \quad 
j=(2i+1)\, \mod N \\
           0 & \quad \mbox{otherwise}
                    \end{array} \right.  ,
\ee
where the number of vertices $N$ is even (Tanner 2000). Every vertex has 
two incoming and
two outgoing edges and the maximal number of steps to reach every vertex 
from every other vertex is $\log_2 N$. It is easy to see that binary 
graphs of dimension $N = 2^k, k\in \N$, so-called de Bruijn graphs 
(Stanley 1999), have a well defined classical limit as defined in 
section \ref{subsec:thermo} which is the dynamics of the Bernoulli shift map. 
All eigenvalues of the transition matrices in this family are zero except the 
leading eigenvalue $\Lambda_0 = 1$; the spectral gap is infinitely large 
and the return probability $IP(n) = 1$ for all $N = 2^k$. It can 
furthermore be shown that every family of binary graphs with dimensions 
$ N = p\, 2^k$, where $p > 1$ is an odd integer, has a classical limit with
a spectral gap $\Delta = \log 2$ independently of $p$. The decomposition 
(\ref{decomp}) of the USE's is unique and the matrices $\tilde{\vec{U}}$ 
are orthogonal consisting of $N/2$ nested $(2\times 2)$ matrices of the form
\[ \tilde{\vec{u}} = \frac{1}{\sqrt{2}} \left( \begin{array}{rr} 
1 & 1\\
1 & -1 \\
\end{array} \right ). 
\]

\begin{figure}
\centering
\centerline{
         \epsfxsize=10cm
         \epsfbox{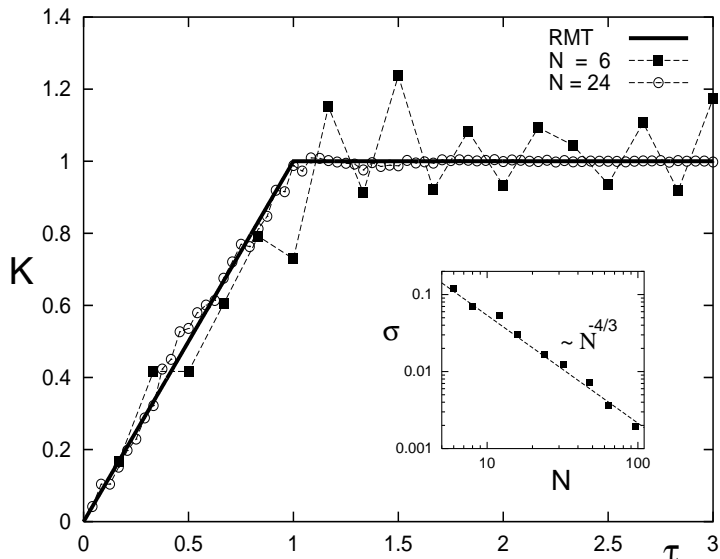}
         }
\caption[]{\small
The form factor for binary graphs with $N = 6$ and $N = 24$; the inset 
shows the standard deviation of $K(\tau)$ from the RMT result as
function of $N$ for $N$ = 6, 8, 12, 16, 24, 32, 48, 64 and 96.
}
\label{Fig:bin}
\end{figure}
The spectral statistics of binary graphs in terms of periodic cycle 
correlations has been studied by Tanner (2000) and convergence 
towards RMT has been found considering transition matrices  up to $N=6$. 
The complexity of the combinatorial formulae describing 
the cycle length correlations made it difficult to advance to larger matrix 
sizes. 
The form factor can be obtained numerically by performing the 
ensemble average (\ref{ava_2}). Fig.\ \ref{Fig:bin} shows the ensemble averaged 
form factor for matrix sizes $N = 6$ and $N=24$, that is, two members of 
the $p = 3$ family. Deviations from the RMT result for the circular 
unitary ensemble (CUE) clearly decrease going from $N=6$ to $N=24$.
The rate of convergence is measured in terms of the 
mean standard deviation $\sigma$ averaged here over the $\tau$ interval 
shown, see the inset of Fig.\ \ref{Fig:bin}. The numerical 
findings indicate a power law behaviour $\sigma(N) \approx N^{-4/3}$ both for 
the $p=1$ and the $p=3$ family, that is, the rate of 
convergence does not depend on the size of the spectral gap. 

\paragraph{Uniformly connected star-graphs.}
\label{uni-star-graphs}
Fully connected graphs with constant transition probabilities, that is, 
\[ t_{ij} = \frac{1}{N} \qquad \mbox{for all} \; i,j = 1, \ldots N \, \] 
display stochastic dynamics with instant complete decay of correlations. 
Every initial probability distribution is mapped onto the equilibrium
state $\tilde{p} = (1/N, \ldots, 1/N)$ in one step and all eigenvalues 
of the transition matrix are zero except for the leading eigenvalue 
$\Lambda_0 = 1$. 

By interpreting the vertices $i$ as the edges of a graph with a single
central vertex, we may view the stochastic process as taking place on 
a star-shaped graph where all transitions between edges through the central
vertex are equally likely (including the `backscattering' processes $i \to i$).
Note that binary graphs of de-Bruijn type with $N = 2^k$ 
are identical to a uniformly connected star-graph after exactly $k$ steps.
The topological entropy $h_t$ measuring the exponential growth 
rate of periodic cycles diverges for star-graphs in the classical limit 
which is in contrast to binary graphs with $h_t = \log 2$. 
\footnote{A diverging topological entropy indicates a singularity in the 
classical dynamics and a star-graph may indeed serve as a model for 
a system with a point-like central scatterer (Berkolaiko \et 2001).}

A possible choice for the matrix $\tilde{\vec{U}}$ in (\ref{decomp})
defining USE's of uniformly connected transition matrices are
symmetric Fourier matrices of the form 
\begin{figure}
\centering
\centerline{
         \epsfxsize=10cm
         \epsfbox{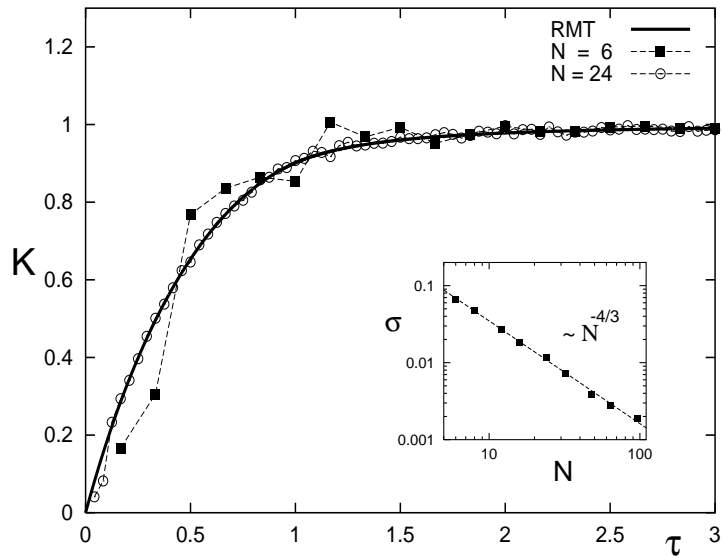}
         }
\caption[]{\small
The form factor for uniformly connected star-graphs and the  
standard deviation $\sigma$.
}
\label{Fig:star}
\end{figure}
\[ \tilde{u}_{nm} = \frac{1}{\sqrt{N}} 
e^{\frac{2 \pi \rm{i}}{N} (n-1)(m-1)} \, .  \]
As a consequence of the symmetry of $\tilde{\vec{U}}$, we expect COE-statistics 
which is indeed observed numerically, see
Fig.\ \ref{Fig:star}. More surprisingly is the fact that  
convergence towards the RMT result is governed by the same power law as for 
binary graphs, see the insets of Fig.\ \ref{Fig:bin} and Fig.\ \ref{Fig:star}. 
One finds numerically $\sigma(N) \approx \frac{3}{4} N^{-4/3}$ for star-graphs,
that is, the standard deviation falls off with the same exponent but a 
slightly smaller prefactor. The rate of convergence is thus insensitive to the 
topology of the graph measured for 
example by the topological entropy or the spectrum of the transition matrix 
as long as a non-zero spectral gap is established.

\subsection{Critical ensembles and deviations from RMT-behavior}
Next, we consider 
two types of systems, namely quantum star-graphs and diffusive networks,
for which the spectral gap closes exactly or faster than the critical rate 
$\Delta \propto 1/N$. The deviations from RMT-statistics have been well 
studied for both types of systems but have not to our knowledge been considered 
in terms of unitary stochastic ensembles and associated transition matrices.

\paragraph{Quantum star-graphs.} 
Quantum star-graphs arise naturally when one quantises a graph with 
a single central vertex attached to $N$ undirected edges. Typical boundary 
conditions imposed on the wave equation at the central vertex 
result in restrictions on the possible transition rates and backscattering 
is greatly favoured. The vertex scattering matrix mentioned in (\ref{S_Smi}), 
which is essentially equivalent to the matrix $\tilde{\vec{U}}$ in 
(\ref{decomp}), is for Neumann boundary conditions of the form
\be{u-star-qu} \tilde{u}_{ij} = - \delta_{ij}  + \frac{2}{N}\; . \ee
The transition matrix corresponding to this orthogonal matrix describes a
Markov process of weakly coupled one dimensional systems with vanishing
coupling strength in the limit $N\to\infty$; the systems is thus 
extended in the sense that it consists of an increasing number of 
almost decoupled equivalent subsystems. 

The Markov processes associated with quantum star-graphs are topologically
equivalent to uniformly connected graphs discussed in \ref{subsec:fingap},
that is, one can move from every vertex to every other vertex in one step.
They do, however, differ greatly in their dynamical properties;
the spectrum of the transition matrix associated with (\ref{u-star-qu})
is highly degenerate and can be given explicitly 
(Kottos and Smilansky 1999), that is,
\[ \lambda_0= 0, \quad \lambda_1, \ldots, \lambda_{N-1} = 
-\log(1 - \frac{4}{N}) \approx \frac{4}{N}\; .\]
Quantum star-graphs have therefore a critical classical spectrum with a 
spectral gap vanishing proportional to $1/N$ and one finds 
spectral statistics intermediate between Poisson and  COE/CUE statistics.
Due to the strong enhancement of backscattering, multiple traversals of the 
period-1 orbits running along the $N$ edges give the dominant 
contributions to the form factor for small  $\tau = \frac{n}{N}$; multiple
repetitions of these orbits are invariant under cyclic permutations
of the symbol code and the diagonal-approximation takes on a form 
differing from Eqn.\ (\ref{D_approx}) (Kottos and Smilansky 1999), i.e.,
\[ K(\tau) \approx \overline{P}(\tau) = e^{-4\tau}  \]
for fixed $\tau$ and $N\to\infty$.  The form factor approaches 1 
for large $\tau$ due to periodic orbit lengths correlations worked out in 
detail by Berkolaiko and Keating (1999), see also Berkolaiko \et (2001).\\

\paragraph{Diffusive networks.} 
The quantum mechanics of classical diffusive systems has been studied mainly 
in the context of Anderson localisation and insulator-metal
(i.e.\ localisation-delocalisation) transitions. A variety of
systems have been considered ranging from disordered conductors
to dynamical localisation in low-dimensional Hamiltonian systems 
as well as various discrete network models, see Dittrich (1996)
and Janssen (1998) for recent review articles. 

\begin{figure}
\centering
\centerline{
         \epsfxsize=9cm
         \epsfbox{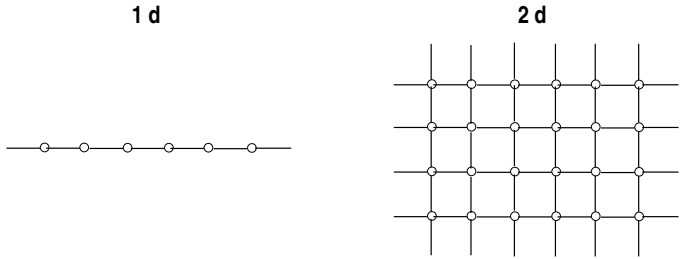}
         }
\caption[]{\small
Diffusive network models in 1 and 2 dimensions where
opposite sides are identified.}
\label{Fig:diffgraph}
\end{figure}
The network models considered here are similar to those studied by 
Shapiro (1982) consisting of regular lattices as shown in Fig.\ 
\ref{Fig:diffgraph}. Each vertex is connected to $2 d$ neighbouring 
vertices via undirected edges and $d$ is identified with the dimension 
of the system. We choose periodic boundary conditions and 
consider cubical networks having the same number of vertices, $L$, along 
each coordinate axis. The total number of vertices is thus $L^d$, the number of
edges equals $2\, d\, L^d$.

We consider a classical Markov process on these networks given
by a transition matrix $\vec{T}_d$ with constant transition probabilities 
\[t_{ij} = \frac{1}{2 d}  \qquad \mbox{if $i$ and $j$ are connected} \]
and $L$ is even. The stochastic dynamics on these networks is in the continuum 
limit $L\to \infty$ in appropriately rescaled units equivalent to 
$d$-dimensional diffusion governed by the diffusion equation 
\be{diff} 
\left(\frac{\partial} {\partial t} - D \vec{\nabla}^2\right) 
\rho(\vec{x},t)= 0
\ee
with diffusion constant $D = \frac{1}{2 d}$. $\rho(x)$ is the continuum limit
of the discrete probability distributions $p$. The low lying eigenvalues
$\lambda$ of $\vec{T}_d$ can be recovered from solving 
 (\ref{diff}) with periodic boundary conditions. The 
eigen-spectrum of (\ref{diff}) is given by 
\be{spec-diff} 
\omega_{\vec{m}} = - \frac{4 \pi^2 D}{L^2} \sum_{i=1}^d m_i^2, 
\ee
where $\vec{m}$ is a $d$-dimensional integer lattice vector.
The eigenvalues of $\vec{T}_d$ converge to the $\{\omega\}$ spectrum
in the large wavelength limit $m_i \ll L$ for all $i$.
Writing the mean return probability (\ref{RP_def}) of the 
Markov process in terms of the spectrum (\ref{spec-diff}), one obtains
\begin{eqnarray}
 \overline{P}(n) &=& \frac{1}{L^d} \Tr \vec{T}_d^n 
\approx \frac{1}{L^d} 
\sum_{\vec{m}} 
\exp\left(-\frac{4 \pi^2 D n}{L^2} \sum_{i=1}^d m_i^2\right)\\
&=& \frac{1}{(4 \pi D n)^{d/2}} 
\sum_{\vec{k}} 
\exp\left(-\frac{L^2}{4 D n}  \sum_{i=1}^d k_i^2\right),
\label{RP_diff}
\end{eqnarray}
where the sums are over all lattice vectors $\vec{m}$, $\vec{l}$, respectively,
and the last equation is obtained by Poisson-summation. The
return probability approaches 
\[ \overline{P}(n) = (4 \pi n D)^{-d/2} = 
\left(\frac{d}{2 \pi n}\right)^{d/2} \]
for large $L$ and $n < L^2$ and converges to $\overline{P}(n) = 1/L^d$ in the 
limit $n\to\infty$ for fixed $L$. The system is thus extended having a well 
defined classical limit which is of course nothing but the diffusion process 
(\ref{diff}) in an infinite domain. 
\begin{figure}
\centering
\centerline{
         \epsfxsize=10cm
         \epsfbox{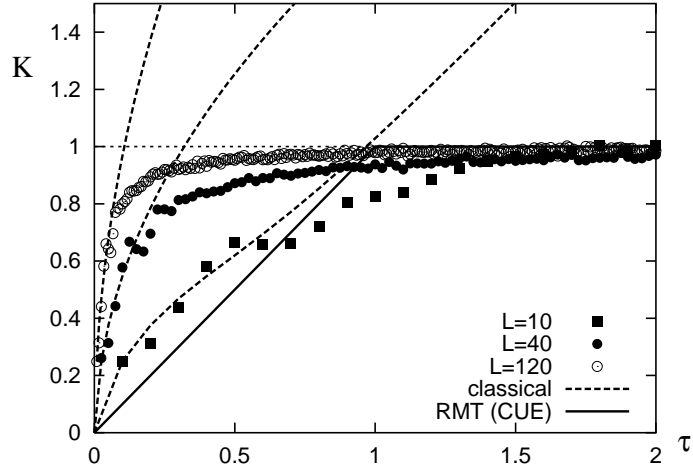}
         }
\caption[]{\small
The form factor for one dimensional diffusive quantum networks of different
size; the data are compared with $n \overline{P}(n)/2d$ 
(dashed line). 
}
\label{Fig:diff-1d}
\end{figure}

\begin{figure}
\centering
\centerline{
         \epsfxsize=10cm
         \epsfbox{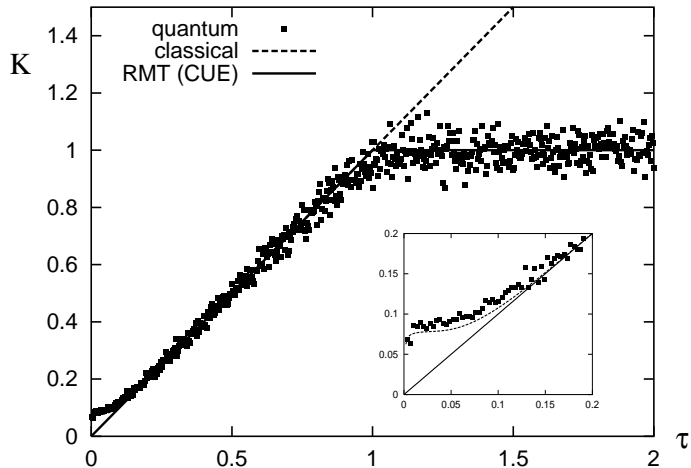}
         }
\caption[]{\small
The quantum form factor for a two-dimensional network of side-length 
$L = 12$ 
compared with classical data (dashed line); a plateau forms for small
$\tau$ at $K = (4 \pi)^{-1} = 0.0795\ldots$, see also inset.
}
\label{Fig:diff-2d}
\end{figure}

The transition matrices $\vec{T}_d$ corresponding to a network of vertices 
on a regular lattice as shown in Fig.\ \ref{Fig:diffgraph} are not unitary
stochastic. USE's on the network can be constructed when considering 
the Markov process describing transitions between adjacent edges of the 
network instead. Edge-edge transition probabilities are again chosen to be
$\tilde{t}_{ij} = 
1/2d$ if $i$ is an an incoming and $j$ an outgoing edge meeting at the same 
vertex. The transition matrix $\tilde{\vec{T}}_d$ of dimension 
$N = 2\, d\, L^d$ is unitary stochastic; it is furthermore easy to see that 
the nonzero eigenvalues of $\tilde{\vec{T}}_d$ coincide with the eigenvalues 
of $\vec{T}_d$, that is 
\[ 
\Tr\vec{T}_d^n = \Tr\tilde{\vec{T}}_d^n  \quad \mbox{for all} \quad n\, .
\]
The matrix $\tilde{\vec{T}}_d$ consists of $L$ coupled 
$2 d\times 2 d$ - matrices. The corresponding unitary stochastic 
ensemble can be written, for example, in terms of 
$L$ coupled uniformly connected star-graphs with $2d$ edges as
described in \ref{subsec:fingap}. Using the diagonal approximation, 
Eqn.\ (\ref{D_approx}), together with the expression for the 
return probability (\ref{RP_diff}), the form factor can for small $\tau$
be written as 
\footnote{There is yet another minor complication; the transition matrices
introduced are not primitive. The network dynamics decomposes into
two unconnected, identical networks when considering the graph for 
$\vec{T}^2_d$,
that is, the two-step dynamics between next-to-nearest neighbour 
vertices and edges. The figures \ref{Fig:diff-1d} and \ref{Fig:diff-2d}
therefore show $K(\tau) = <\frac{1}{2N} |\Tr 
\vec{U}^{2n}|^2>_{U(N,\vec{T}_d)}$ with $\tau = 2n/N$.}
\be{K_diff}
K(\tau) \approx \frac{n}{2 d L^d}
\Tr \tilde{\vec{T}}^n_d =  n  \frac{L^d}{2 d L^d} \overline{P}(n) 
\approx \frac{1}{2 d} \frac{1}{(4 \pi D )^{d/2}} (N \tau)^{1-d/2} \; 
\ee
where we first made the diagonal approximation valid in the limit $n/L^d \to 0$
and $n \to \infty$ and the second approximation is applicable for $n < L^2$. 
One finds in particular for one dimensional diffusion
\[ K(\tau)  \approx  \frac{1}{2}
\sqrt{\frac{N}{2 \pi} \tau } 
\qquad \mbox{for small } \tau \; .
\]
Quantum interference leads to deviations from the diagonal approximation at 
$\tau \approx \max(\frac{8\pi}{N}, 1)$ (Schanz and Smilansky 2000) at which 
the form factor approaches 1, see Fig.\ \ref{Fig:diff-1d}. Eigenstates of the 
USE are all localised
in the classical limit, level repulsion between eigenvalues vanishes and the 
spectral statistics converges to the Poissonian limit.
The form factor forms a plateau for $d = 2$ and small 
$\tau$ values, that is,
\[ K(\tau)_{U(N,\vec{T_2})}  \approx  \frac{1}{4 \pi}
\qquad \mbox{for} \quad \tau < \frac{1}{4 \pi}\, ,\]
which persists in the limit $N\to\infty$, see Fig.\ \ref{Fig:diff-2d}. We thus
find a stationary distribution which converges neither to the CUE nor to 
the Poisson - limit which is a clear indication for an ensemble being critical.
Finally, considering $d >2$, we expect convergence of the form factor to the
CUE result for large $N$ which is confirmed by numerical calculations (not 
shown here).

The influence of the dimension $d$ on the small $\tau$ behaviour of $K(\tau)$ 
in diffusive systems has been described in detail  by, for example,
Dittrich (1996) and references therein. It is interesting to reconsider
these results in terms of the spectral gap as described in section 
\ref{subsec:thermo}. The spectral gap between the leading and next to leading 
eigenvalues
of $\vec{T}_d$ (and thus of $\tilde{\vec{T}}_d$) can be read off from 
Eqn.\ (\ref{spec-diff}); one obtains 
\[\Delta_N = \frac{4 \pi^2 D}{L^2} = \frac{4 \pi^2 D (2 d)^{2/d}}{N^{2/d}} \]
that is, the spectral gap falls off slower than $1/N$ for $d>2$ only!
The case $d = 2$ is critical and deviations from RMT indeed remain stationary
in the classical limit. One-dimensional diffusion is in this sense
super-critical leading to Poisson statistics.
It is worthwhile noting that the Kolmogorov-Sinai entropy for the lattices
considered is positive for all $d$ including $d=1$ and 2, that is, 
$K_{KS} = \log 2d \, >0$. A positive KS-entropy does therefore not 
necessarily imply RMT-statistics.

\section{Conclusion}
We propose a new way to partition the unitary group into ensembles of 
unitary matrices specified in terms of unitary-stochastic transition matrices.
This provides a framework to study systematically the connections 
between eigenvalue statistics of unitary matrices and the properties of an 
associated classical dynamics, here the stochastic dynamics on a graph. We 
define a classical limit of a family of stochastic 
networks with increasing network size. This makes it possible to 
give a strict criterion distinguishing between unitary stochastic 
ensembles whose spectral statistics converges towards the standard RMT results 
for $N\to\infty$ and those whose statistics does not. Many questions, 
however, remain unanswered. The most interesting one is certainly 
how to link the universal properties of spectra of unitary matrices found 
on all scales (and not only for small $\tau$ or for long-range 
correlations) to properties of a classical dynamics. Universality 
suggests a common principle; the attempts made so far to describe spectral 
properties of graphs beyond the 
diagonal approximation do, however, all rely heavily on the specific system 
under consideration (Kottos and Smilansky 1999, Berkolaiko and Keating 
1999, Schanz and Smilansky 1999, 2000, Tanner 2000) and a general scheme 
is not yet in sight. One might furthermore expect that the spectral properties 
of almost all matrices within  an ensemble (after applying local averaging 
within a given spectrum) coincide with the ensemble average, an assumption 
which remains to be shown to be true.  Finally, it would be interesting to 
study general quantum systems corresponding to a classical deterministic 
dynamics in terms of graphs. Connections between the semiclassical 
limit of a series of USE's and the semiclassical limit of a quantum map 
might be a way forward to understand universality in quantum spectra in 
general. \\

\noindent
{\large \bf Acknowledgments}\\

\noindent
I would like to thank Prot Pako\'nski, Uzy Smilansky and Robert Whitney 
for stimulating discussions and Uzy Smilansky for having been such a 
good host at the Weizmann 
Institute, where parts of this work have been carried out. I thank Stephen 
Creagh and Jon Keating for valuable comments and for carefully reading the 
manuscript. I am also grateful for the hospitality experienced during
numerous stays at BRIMS, Hewlett--Packard Laboratories in Bristol and for
financial support received from the Nuffield Foundation.

\end{document}